\documentclass{aa}

\usepackage{graphicx}
\usepackage{txfonts}

\usepackage{color,xcolor,xspace,pdflscape,dblfloatfix,caption}
\usepackage{hyperref} 

\newcommand{\equ}[1]{Eq.~\ref{eq:#1}}

\newcommand{\fig}[1]{Fig.~\ref{fig:#1}}

\newcommand{\tab}[1]{Table~\ref{tab:#1}}

\newcommand{\sect}[1]{Sect.~\ref{sec:#1}}

\newcommand{\lcdm}[0]{$\Lambda$CDM\xspace}

\usepackage{array}

\begin{document}

\title{Deviations from the radial acceleration relation in the central galaxies of clusters, subclusters, and groups}

\authorrunning{Michal B\'ilek et al.}

   \author{
   Michal B\'ilek \inst{1},
   Florent Renaud\inst{2,3},
   Srdjan Samurovi\'c\inst{1}
   }

   \institute{Astronomical Observatory of Belgrade, Volgina 7, 11060 Belgrade, Serbia,     \email{michal.bilek@aob.rs}
    \and Observatoire Astronomique de Strasbourg, Universit\'e de Strasbourg, CNRS UMR 7550, F-67000 Strasbourg, France
    \and University of Strasbourg Institute for Advanced Study, 5 all\'ee du G\'en\'eral Rouvillois, F-67083 Strasbourg, France
             }
   \date{Received: ...; accepted ...}

 \abstract 
 {
{Most galaxies closely follow the radial acceleration relation (RAR), which tightly links the observed accelerations to those predicted by Newtonian gravity from visible baryonic matter. Galaxy clusters, however, deviate from this relation. Several explanations have been proposed. Some of them predict that even some galaxies in clusters should deviate, but this hypothesis remains largely untested. We test it here by analyzing acceleration profiles for 17 early-type galaxies, derived from Jeans modeling of their globular cluster systems in our older work. Our sample spans central galaxies in clusters and groups, noncentral galaxies, isolated ones, and—uniquely for this paper—centrals in galactic subclusters, which are smaller clusters being accreted by larger ones. We compare these profiles to the standard RAR for noncluster galaxies and its counterpart for clusters. We find that isolated and noncentral galaxies adhere to the standard RAR. In contrast, central galaxies of clusters, subclusters, and groups { exhibit enhanced accelerations in most cases, tracing  instead }the cluster acceleration behavior either partly or fully. The radius at which divergence from the standard RAR begins tends to decrease with increasing group mass. These findings imply that if cluster fields depart from the standard RAR due to undetected {material, it} must be dynamically cold and collisionless, such as nonbaryonic cold dark matter, but also compact clouds of cold gas.}       
 }

               
   \maketitle
\section{Introduction}

Most galaxies follow the radial acceleration relation (RAR)\citep{mcgaugh16,lelli17,mistele24}, which connects the observed acceleration in galaxies $a$ to the distribution of luminous matter. Namely, for a Newtonian gravitational acceleration $a_N$ caused by baryons, the measured acceleration is very well modeled by \citep{mcgaugh16}
\begin{equation}
    a = \frac{a_N}{1-\exp(-\sqrt{a_N/a_0})},
    \label{eq:rar}
\end{equation}
where $a_0 = 1.2\times10^{-10}\,$m$\,$s$^{-2}$ is the so-called galactic acceleration scale. The intrinsic scatter of this relation is extremely small, at least for late type galaxies \citep{li18,stiskalek23,desmond23,varasteanu25}. The relation is well established observationally, but was first predicted theoretically in the framework of modified Newtonian dynamics (MOND; \citealp{milg83a}).

{In lambda cold dark matter cosmology (\lcdm), the RAR implies that baryons and dark matter are always spatially distributed to satisfy \equ{rar}. However, this is not expected a priori. In order to explain why the baryonic fractions of galaxies are below the cosmic mean, one must invoke baryonic outflows. They are supposed to be caused by mechanisms such as winds from stars, supernova explosions, active galactic nuclei, or reionization. The dominant mechanism is believed to strongly depend on galaxy mass \citep{hopkins14,penoyre17,wechsler18,harrison18,remus22,sales22,contini25}. Also, the fraction of galaxy material assembled by smooth gas accretion of the intergalactic medium and mergers varies from galaxy to galaxy, correlating with mass. It is therefore surprising that all galaxies of all masses follow the same RAR.  Simulations and analytical models have been used to investigate RAR  expectations using the \lcdm framework \citep{wu15,dicintio16,keller17,ludlow17,tenneti18,brouwer21,li22}. They reproduce the observed RAR to varying extents, leaving the prediction unclear. Improper comparisons with the data have also been noted \citep{milg02b,milg16,li22}. Thus, it remains uncertain whether \lcdm can explain the RAR, spurring investigations of alternatives that guarantee it. Some of them assign special properties to dark matter particles \citep{blanchet08,zhao08,berezhiani15,famaey20}, and others are based on modifications of gravity or inertia \citep{moffat06,verlinde17}, whose best-known representative is MOND \citep{milg83a,famaey12}. }

However, in galaxy clusters and {some galaxy groups}, the observed accelerations lie above the RAR predictions {\citep{the88,sanders94,sanders99,sanders03,milgcbdm,angus08,ettori19,li24,kelleher24,famaey25, mistele25}}. Several ideas are proposed to reconcile this discrepancy with MOND and similar theories that predict the RAR universally. One direction involves modifications of such theories. For example, for MOND it was proposed that its acceleration scale $a_0$ depends on the local value of the gravitational potential (\citealp{zhao12}; see also, \citealp{gqumond,durakovic24}). Another idea keeps the original equations and constants, but postulates the existence of additional nonluminous or faint matter in galaxy clusters (see a review in \citealp{kelleher24}). This matter can be hot, such as sterile neutrinos \citep{sanders03,angus10} with velocities too high to allow them to cluster into low-mass halos around isolated galaxies, but compatible with the formation of massive halos in galaxy clusters. The last of the most discussed alternatives invokes the presence of cold compact baryonic objects in galaxy clusters, which do not emit or absorb much radiation \citep{milgcbdm, kelleher24}. The best candidates are compact clouds of molecular gas. There are some indirect pieces of evidence for this, some relying on MOND \citep{famaey25}, others not  \citep{fabian22}. 

Importantly, for this paper, some of the above solutions imply that even the central galaxies of clusters {and groups} might show deviations from the RAR. This could eliminate some of the above alternatives{, as we discuss below}. Detailed studies of gravitational fields in the {centrals} remain rare. The few existing studies, in fact,  suggest that the {centrals of the clusters} deviate from the RAR \citep{richtler08,bil19,hilker18, tian20,tian24}. 

{Several observational studies have been conducted of the relation between $a$ and $a_N$ in clusters \citep{tian20,chan20,pradyumna21,eckert22,liu23,li23,kelleher24,tian24,mistele25}. They obtained generally comparable results. We are aware of only one work that compared the different results in detail, namely \citet{pradyumna21}, one of the earliest works, which found that the results are not entirely consistent with each other, indicating systematic errors. The influence of systematics has also been demonstrated in \citet{li23,kelleher24,mistele25}. For concreteness, here we will compare our galaxies to the cluster $a$ and $a_N$ relation by \citet{tian24}, which is a development of the work \citet{tian20}.} 

{In particular, \citet{tian20}  found that in galaxy clusters, down to the brightest cluster galaxy (BCG) region, there is a correlation between $a$ and $a_N$, similar to the standard RAR but with an acceleration scale approximately ten times higher. These measurements were taken at distances of hundreds of kiloparsecs from the cluster centers and at about one effective radius of the BCGs. Later, \citet{tian24} confirmed that BCGs follow this correlation when they derive their gravitational fields at one effective radius from stellar kinematics.} The scatter of the relation combining both types of objects (including observational uncertainties) is about 0.1 \,dex. 

In \citet{bil19} (BSR19 hereafter), we derived the acceleration profiles of 17 early-type galaxies from the kinematics of their globular cluster (GC) systems. Notably, they were published before all the $a$--$a_N$ studies in clusters. In the present work, we confront these profiles with the relation {of Tian et al.  } Our sample comprises all central and noncentral galaxies of clusters, groups, and isolated galaxies. Additionally, here we pay attention to the central galaxies of subclusters, which are groups and clusters currently undergoing a merger with a more massive cluster. This provides a variety of environments in which we can study how well galaxies follow the RAR.

This paper is organized as follows. { Section~\ref{sec:method} presents the sample of investigated galaxies and summarizes the method BSR19 used to derive the radial acceleration profiles. Section~\ref{sec:env} describes the environments of the galaxies and their place in them.} In \sect{res}, we show that {the central galaxies of their structures} tend to deviate from the RAR. We conclude {in \sect{conc}} that the most natural interpretation is that deviations from the RAR are caused by the presence of collisionless, dynamically cold, invisible material in clusters.
  
\hspace*{-0.5cm}
\begin{table}
\caption{Environmental classification of the galaxies. }
\label{tab:env}
\centering
\begin{tabular}{@{}>{\raggedright\arraybackslash}p{0.4cm} >{\raggedright\arraybackslash}p{1.4cm} >{\raggedright\arraybackslash}p{5.1cm}>{\raggedright\arraybackslash}p{0.8cm}@{}}
\hline
Rank & Galaxy & Environment & $r_\mathrm{max}$  \\
 & & & [kpc]\\
\hline\hline
1 & NGC\,4486 & Central of Virgo A, $5 \times 10^{14}\,M_\sun^\mathrm{(1)}$  & 140 \\
2 & NGC\,4472 & Central of Virgo B, $1 \times 10^{14}\,M_\sun^\mathrm{(1)}$ & 45 \\
3 & NGC\,1399 & Central of Fornax Cluster, $9 \times 10^{13}\,M_\sun^\mathrm{(2)}$ & 106 \\
4 & NGC\,5846 & Central of its group, $8 \times 10^{13}\,M_\sun^\mathrm{(2)}$ & 64 \\
5 & NGC\,1407 & Central of  its group, $6 \times 10^{13}\,M_\sun^\mathrm{(2)}$ & 122 \\
6 & NGC\,4365 & Central of Virgo W' cloud, $3 \times 10^{13}\,M_\sun^\mathrm{(1)}$ & 69\\
7 & NGC\,4649 & Central of Virgo C, $3 \times 10^{13}\,M_\sun^\mathrm{(1)}$ & 105\\
8 & NGC\,5128 & Central of its group $8 \times 10^{12}\,M_\sun^\mathrm{(2)}$  & 52\\
9 & NGC\,4278 & Central of a small localized clump in the loose extended Coma~I cloud$^\mathrm{(3)}$ & 39\\
10 & NGC\,1023 & Central of its group$^\mathrm{(2)}$, but visually appears virtually isolated  & 26\\
11 & NGC\,2768 & Isolated or group member$^\mathrm{(4)}$  & 60\\
12 & NGC\,3115 & Isolated$^\mathrm{(5)}$ & 22 \\
13 & NGC\,0821 & Isolated$^\mathrm{(6)}$ & 35 \\
14n & NGC\,4494 & Non-central of the loose Coma~II cloud$^\mathrm{(3)}$  & 38\\
15n & NGC\,3377 & Non-central of Leo~I group$^\mathrm{(3,7)}$  & 31\\
16n & NGC\,1400 & Non-central of NGC\,1407 group$^\mathrm{(2)}$ &  71\\
17n & NGC\,4526 & Non-central of Virgo B$^\mathrm{(8)}$ & 33 \\
\hline
\end{tabular}
\tablefoot{A lower rank means the galaxy dominates a larger structure. The suffix ``n'' of the rank indicates noncentral galaxies.   The  mass stands for the virial mass. {The $r_\mathrm{max}$ column indicates the galactocentric distance of the last bin of measurements of the gravitational field.}}
\tablebib{
(1)~\sect{env}; (2) \citet{tully15}; (3) \citet{ferrarese00}; (4) \citet{forbes12}; (5) \citet{karachentsev22}; (6) \citet{spitler08}; (7) \citet{muller18b}; (8) \citet{vattakunnel10}.
}
\end{table}

\section{{Galaxy sample and the derivation of their profiles of gravitational acceleration}}
\label{sec:method}
{The 17 elliptical and lenticular galaxies analyzed here are listed in \tab{env}. It is the same sample as in BSR19  but without the simulated galaxies. The stellar masses of all included galaxies are comparable, ranging from about $10^{10}$ to $10^{11}\,M_\sun$. The $r_\mathrm{max}$ column  indicates the greatest radius of the measurement of the gravitational field. Many more properties of the galaxies were listed in Tables~1 and~2 in BSR19.

Here we briefly summarize how the acceleration profiles were derived in BSR19{ (see Sect.~7 of that paper for details). Jeans modeling of the radial velocities of GCs was done.} The gravitational fields were parametrized by a sum of the Newtonian gravitational potential resulting from the stars of the galaxy and a Navarro-Frenk-White profile \citep{nfw} halo. The free parameters are the mass-to-light ratio of the stellar component and the scale radius and concentration of the halo. There are no restrictions or priors other than that the last two parameters must be positive and that the total gravitational force must be attractive {at any radius}. This approach can model a wide diversity of acceleration profiles. We checked that our models match the MOND predictions  within a few percent.
For each galaxy, three fits of the gravitational profile were made, differing by the assumed velocity dispersion profiles: the isotropic and tangential models have a constant anisotropy of zero or -0.5, respectively. The externally radial (called also the ``literature'') model derived from cosmological simulations has an anisotropy parameter $\beta$ increasing from zero in the galaxy center to 0.5 at infinity. This method has been validated against a \lcdm zoom-in cosmological simulation from \citet{Renaud17}. {The resulting best-fit parameters of the models were not published in BSR19, and therefore we publish them here in \tab{bestfit}.} 

\begin{figure*}
  \includegraphics[width=17cm]{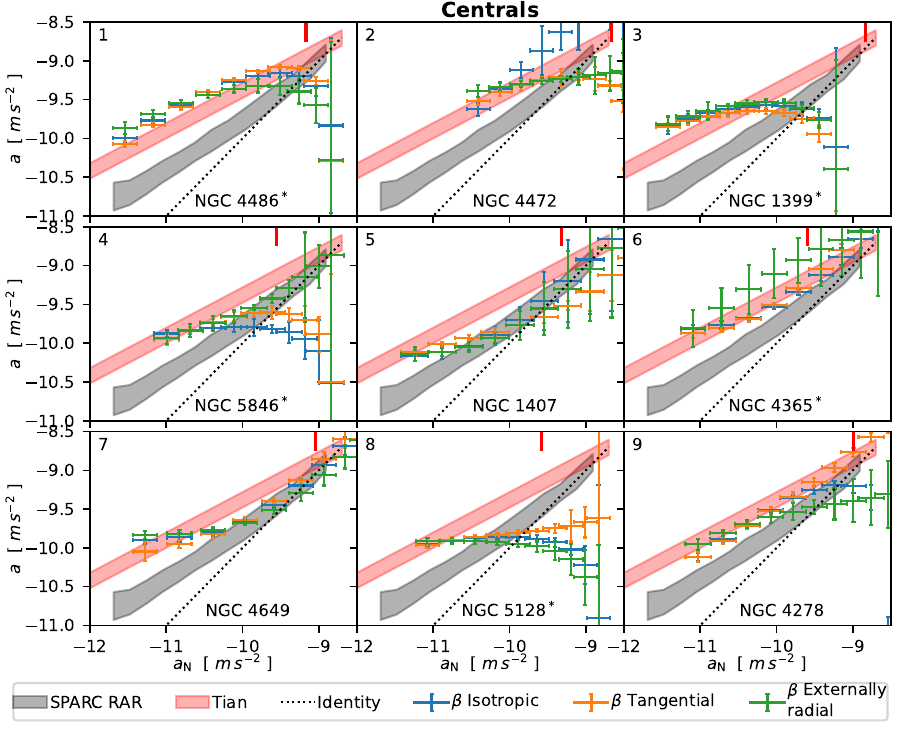}
     \caption{RAR for the central galaxies of clusters, subclusters, and groups. For each galaxy, three estimates of the net acceleration are given, depending on the anisotropy parameter (see \sect{method}). The error bars comprise all uncertainty in galaxy distances, stellar mass-to-light ratios, and uncertainties from the number of dynamical tracers employed. They are compared to the RAR of spiral galaxies from the SPARC sample \citep{mcgaugh16}, and to the relation found by \citet{tian24} for central galaxies of clusters. Shaded areas show the $1\sigma$ measured scatter. The numbers in the top-left corners indicate the rank of the galaxy in \tab{env}. Galaxies with asterisks in their name show signs of ongoing interactions. The short vertical red bar in each panel indicates one effective radius of the galaxy (BSR19).}
     \label{fig:centrals}
\end{figure*}

\begin{figure*}
  \includegraphics[width=17cm]{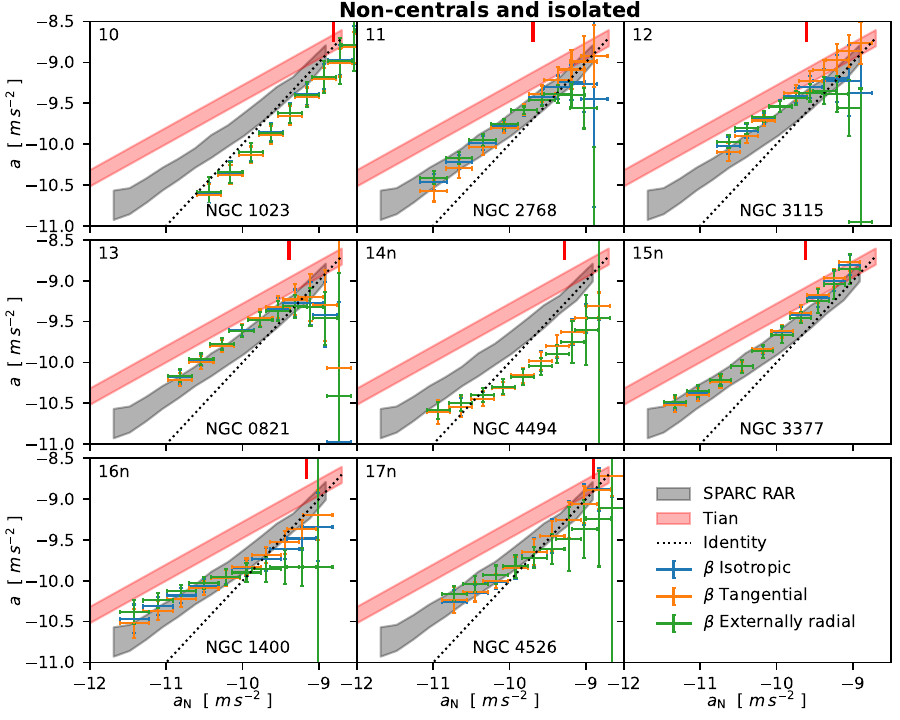}
     \caption{ Same as \fig{centrals} but for the noncentral and isolated galaxies. }
     \label{fig:noncentrals}
\end{figure*}

\section{{Galaxy environmental status}}
\label{sec:env}
The galaxies' environmental status has mostly been compiled from the literature. We also visually inspected the appearance of the neighborhood of galaxies in all sky imaging surveys in Aladin \citep{aladin}. The total environmental status of the galaxies investigated is described in the {third} column of \tab{env}. Here we have additional remarks.

\begin{itemize}
    \item The Virgo Cluster consists of several subclusters \citep[e.g.,][]{cantiello24}. Our sample includes the centrals of the three main subclusters of Virgo called A, B, and C, and of the infalling galaxy group called Cloud W'. We could not find virial masses in the literature. We estimated them ourselves from the virial relation $M_v = KR\sigma^2/G$ \citep{tully15}. Here $R$ stands for a specific type of weighed average of the individual cluster-centric distances of the galaxies, but it was not available.  We instead used  the total clump radius and velocity dispersion $\sigma$ given in Tab.~1 of \citet{morgan25}. We empirically assessed the value of $K$ such that the virial masses of clumps A, B, and C add to the well-established virial mass of the entire Virgo Cluster ($6.3\times10^{14}\,M_\sun$ \citealt{kashibadze20}). The results are in \tab{env}.
    \item NGC\,1023 is classified as the central of its own small group \citep{tully15}. However, according to our visual inspection, it appears almost isolated, with only much less luminous galaxies  around it.
    \item The Coma~I and Coma-II Clouds lie in the foreground of the Coma Cluster, but are not related to it. The distribution of galaxies in this region is messy with poorly defined structures \citep{gregory77, ferrarese00}. There is a localized overdensity around NGC\,4278 \citep{ferrarese00, bosseli09}. It is less dense than the outskirts of the Virgo Cluster \citep{bosseli09}. The Coma~II cloud hosts NGC\,4494, but this galaxy does not occupy any central position \citep{ferrarese00}.
\end{itemize}

In MOND, when considering galaxy clusters as wholes, the highest ratio of invisible to visible matter is encountered in the centers of the clusters \citep{sanders03,milgcbdm}.  From this we can expect that galaxies that are the central object of their cluster, subcluster, or group are the most likely to {deviate from the RAR}. Also, small groups need little or none of the invisible material \citep{famaey12,milg18,milg19}.  From this we can expect that deviations of the central galaxies from the standard RAR scale with the mass of the host structure.

We sorted galaxies in our sample by deviation from the RAR we expected and assigned them the rank shown in \tab{env}, starting with galaxies whose environments have the highest virial masses, i.e., centrals of massive clusters. Galaxies hosting no, or only faint, satellites, which thus appear as virtually isolated, follow. Finally, we add noncentral galaxies that do not dominate their environment, starting with galaxies that are not dominant in loose environments, and then noncentral galaxies of massive clusters.
The ordering of the noncentral galaxies is somewhat arbitrary, but as we see at the end, it is not important.  For clarity, the rank numbers of the noncentral galaxies were given the suffix ``n''.

{One might wonder whether our galaxies are BCGs, for example, for comparison with the literature results that investigate the RAR in BCGs, such as \citet{tian24}. The central galaxies are often BCGs, but not always. To determine if our galaxies are BCGs, we consulted the group and cluster catalog of \citet{tempel16}. Specifically, we determined the parent structure for each galaxy and verified whether it is the brightest member based on extinction-corrected $K_s$-band magnitude listed in the catalog. Our central and isolated galaxies all qualify as BCGs in a broader sense, whereas noncentral galaxies qualify as non-BCGs.

We define a BCG in a ``broader sense'' as the most luminous galaxy within its local environment, irrespective of group or cluster conventions---a distinction that is semantic rather than physical bimodality. Even cataloged isolated galaxies typically harbor satellites, forming de facto groups. The Virgo Cluster, treated as a single entity, harbors two BCGs of virtually identical brightness: NGC\,4486 ($K_s = 5.90$\,mag) and NGC\,4649 ($K_s = 5.82$\,mag). Similarly, NGC\,4278 ($K_s = 7.26$\,mag) is practically as bright as the brightest member of the group ($K_s = 7.12$\,mag). The catalog of \citet{tempel16} classifies NGC\,1400 as isolated, but \citet{tully15} as a member of the NGC\,1407 group, and explains the high relative velocity by an ongoing pericentric passage. We adopt the latter classification because of its basis in a detailed manual analysis.

}

\section{Results}
\label{sec:res}
We split the sample into the eight least dominant and the nine most dominant galaxies, according to their rank. Figures~\ref{fig:centrals} and~\ref{fig:noncentrals} show the relation between $a_\mathrm{N}$ and $a$ for the dominant and nondominant subsamples, respectively. {This is compared to the standard RAR and the  $a_\mathrm{N}$-$a$ relation by \citet{tian24}.} The difference between the dominant and nondominant galaxies is obvious. The isolated and noncentral cases generally tend to follow the standard RAR. In contrast, the central galaxies of clusters, subclusters, and groups tend to deviate. Some of these galaxies follow the relation by \citet{tian24} at all radii, others stick to the standard RAR in the center and switch to the Tian et al. relation further away. There is a tendency for the more dominant galaxies to switch to the relation by \citet{tian24} closer to the center. At the furthest data point (that is, the lowest acceleration), all galaxies in \fig{centrals} approximately agree with this relation.

\section{Discussion and conclusions}
\label{sec:conc}

First, it is useful to comment on the limitations of the results. In \fig{noncentrals}, there are two galaxies{, NGC\,1023 and NGC\,4494,} whose reconstructed gravitational field is even lower than that generated by Newtonian gravity and baryons alone. They are still consistent with the RAR within about $2\sigma$ error bars, but it is possible that the deviation is caused by the undetected rotation of their GC systems in the plane of sky. The procedure used in BSR19 accounts only for the line-of-sight component of rotation. {Other possible sources of systematics are incorrectly determined stellar masses and distances of the galaxies}. It was noted in BSR19 that the galaxies marked by the asterisks in the figures show signs of interactions, which could bias the estimates of the gravitational fields. In \fig{centrals}, we can note that all galaxies in which observed acceleration falls below the RAR prediction in the center are interacting. Actually, in BSR19 it was proposed that the interactions cause the disagreement between the inferred acceleration profiles from MOND and the RAR, but the magnitude of these effects has never been investigated by simulations. Having warned about the potential systematics, we now discuss our results at face value.

Hot dark matter, such as sterile neutrinos \citep{angus10}, has been proposed to explain the deviation of galaxy clusters from the RAR. These dynamically hot particles are not expected to cluster in individual galaxies. Yet we see deviations from the RAR in them. Our findings thus disfavor hot dark matter as a reason for the deviation of the clusters from the RAR. Dedicated calculations are necessary to determine which dark particle speeds are excluded.

The upward deviation from the RAR has already been reported for a few brightest cluster galaxies \citep{richtler08, bil19, hilker18,tian24}. In the present work, we extend this conclusion to the centrals of subclusters. If the deviations from the RAR are caused  by undetected baryonic matter,  our study suggests that it cannot be of a collisional nature, as ram pressure stripping would remove such material from the centrals of subclusters. However, the centrals of the Virgo subclusters A, B, and C are around 1\,Mpc from each other. If they are approaching each other for the first time, ram pressure stripping might not yet have a significant influence at such separations. The collisionless nature of the missing material is more strongly supported by observations of colliding systems similar to the Bullet Cluster \citep{clowe06, bradac08,harvey15,finner23,finner25}. We note that there was one well-known counterexample, the Train Wreck Cluster, where the visible and dark matter did not separate in the collision  \citep{mahdavi07}. However, no other objects like this have been found and subsequent studies of the Train Wreck Cluster reduced tension with collisionless dark matter \citep{clowe12,peel17}. The displacement of the gravitational centers from the baryonic centers in colliding galaxy clusters also disagrees with the original nonrelativistic EMOND modification of MOND \citep{zhao12}. 

{We conclude that in the theories that guarantee the RAR, the deviations of clusters, groups, and their central galaxies are best explained by collisionless dynamically cold invisible material. This can be either the nonbaryonic dark matter known from the \lcdm model, but also compact baryonic objects \citep{milgcbdm}, such as clouds of the cold phase of the intercluster medium \citep{kelleher24,famaey25}. Independent astrophysical evidence for the existence of such clouds has been observed \citep{fabian22}. }

\begin{acknowledgements}
We thank the anonymous referee for constructive comments that significantly improved the manuscript.
This research was supported by the Ministry of Science, Technological Development and Innovation of the Republic of Serbia under contract no.  451-03-33/2026-03/200002 with the Astronomical Observatory of Belgrade. 

\end{acknowledgements}

\bibliographystyle{aa}
\bibliography{literature}

\begin{appendix}
\onecolumn
\section{Best fit profile parameters}
In this section, we provide \tab{bestfit} of the best-fit NFW parameters of the kinematics of the GC systems obtained in Sect.~7 of BSR19. We remind that they purely serve for parametrization of the gravitational field and therefore the parameters can be nonphysical, such as negative  stellar mass-to-light ratios. For physically meaningful NFW models, see BSR19. For each galaxy, there are three models corresponding to three different assumptions on the anisotropy parameter $\beta$, as explained in \sect{method}: isotropic (``iso''), negative (``neg''), and literature (``lit''). For each of them, the table gives the best fit of the stellar mass-to-light ratio $M/L$, and the virial mass $M_v$ and  scale radius $r_s$ of the NFW halo.

\begin{table*}[h!]
\caption{Best fit profile parameters.}  
\label{tab:bestfit} 
\begin{tabular}{p{1.4cm} |p{1.2cm}p{1.2cm}p{1.2cm} |p{1.2cm}p{1.2cm}p{1.2cm}| p{1.2cm}p{1.2cm}p{1.2cm}}
\hline\hline
 & & $\beta_\mathrm{iso}$ & & & $\beta_\mathrm{neg}$ & & & $\beta_\mathrm{lit}$ & \\\cline{2-10}
Name   & $M/L$ & $\log_{10}\frac{M_{v}}{M_\sun}$ &  $\log_{10}\frac{r_{s}}{\mathrm{kpc}}$ & $M/L$ & $\log_{10}\frac{M_{v}}{M_\sun}$ &  $\log_{10}\frac{r_{s}}{\mathrm{kpc}}$ & $M/L$ & $\log_{10}\frac{M_{v}}{M_\sun}$ &  $\log_{10}\frac{r_{s}}{\mathrm{kpc}}$   \\
\hline                                
\csname @@input\endcsname "bestfittab.txt"
\hline      
\end{tabular}
\end{table*}

\end{appendix}
\label{LastPage} 
\end{document}